\documentclass{kluwer}

\usepackage{graphicx}

\begin{document}
\begin{article}

\begin{opening}

\title{A Multi-Phase Chemo-Dynamical SPH Code for Galaxy Evolution.
Testing the Code.}

\author{Peter \surname{Berczik}\email{berczik@mao.kiev.ua}}
\institute{Main Astron.\ Obs.\ of Ukrainian National Academy of
Sciences, Kiev, Ukraine}

\author{Gerhard \surname{Hensler}}
\author{Christian \surname{Theis}}
\institute{Institut f\"ur Theoretische Physik und Astrophysik,
University of Kiel, Germany}

\author{Rainer \surname{Spurzem}}
\institute{Astronomisches Rechen-Institut, Heidelberg, Germany}

\runningauthor{Peter Berczik et al.}
\runningtitle{MP-CD-SPH. Code testing}


\begin{abstract}
In this paper we present some test results of our newly developed
Multi-Phase Chemo-Dynamical Smoothed Particle Hydrodynamics
(MP-CD-SPH) code for galaxy evolution. At first, we present
a test of the ``pure'' hydro SPH part of the code. Then
we describe and test the multi-phase description of the gaseous components
of the interstellar matter. In this second part we also
compare our condensation and evaporation description with the
results of a previous 2d multi-phase hydrodynamic mesh code.
\end{abstract}



\end{opening}


\section{Introduction}

Since several years Smoothed Particle Hydrodynamics (SPH) 
\cite{M1992} has been applied successfully to study the formation and
evolution of galaxies. Its Lagrangian nature as well as its easy
implementation together with standard N-body codes allows for a
simultaneous description of complex dark matter-gas-stellar
compositions \cite{NW1993, MH1996}. Nevertheless, until now the
codes lack processes that are based on the coexistence of
different phases of the interstellar medium (ISM), 
mainly dissipative, dynamical and stellar feedback, 
element distributions, etc. We have therefore
developed a 3d chemo-dynamical code which is based on our single
phase galactic evolutionary program \cite{Ber1999, Ber2000}. This
code includes many complex effects such as a multi-phase ISM,
cloud-cloud collisions, the drag force between different ISM
components, condensation and evaporation of clouds (CE), star
formation (SF) and stellar feedback (FB). The more detailed
description of the new code will be presented in a 
comprehensive paper (Berczik et al., in preparation).

\section{The hydro code test}

\begin{figure}[h]
\tabcapfont
\centerline{%
\begin{tabular}{c@{\hspace{0.2in}}c}
\includegraphics[width=2.2in]{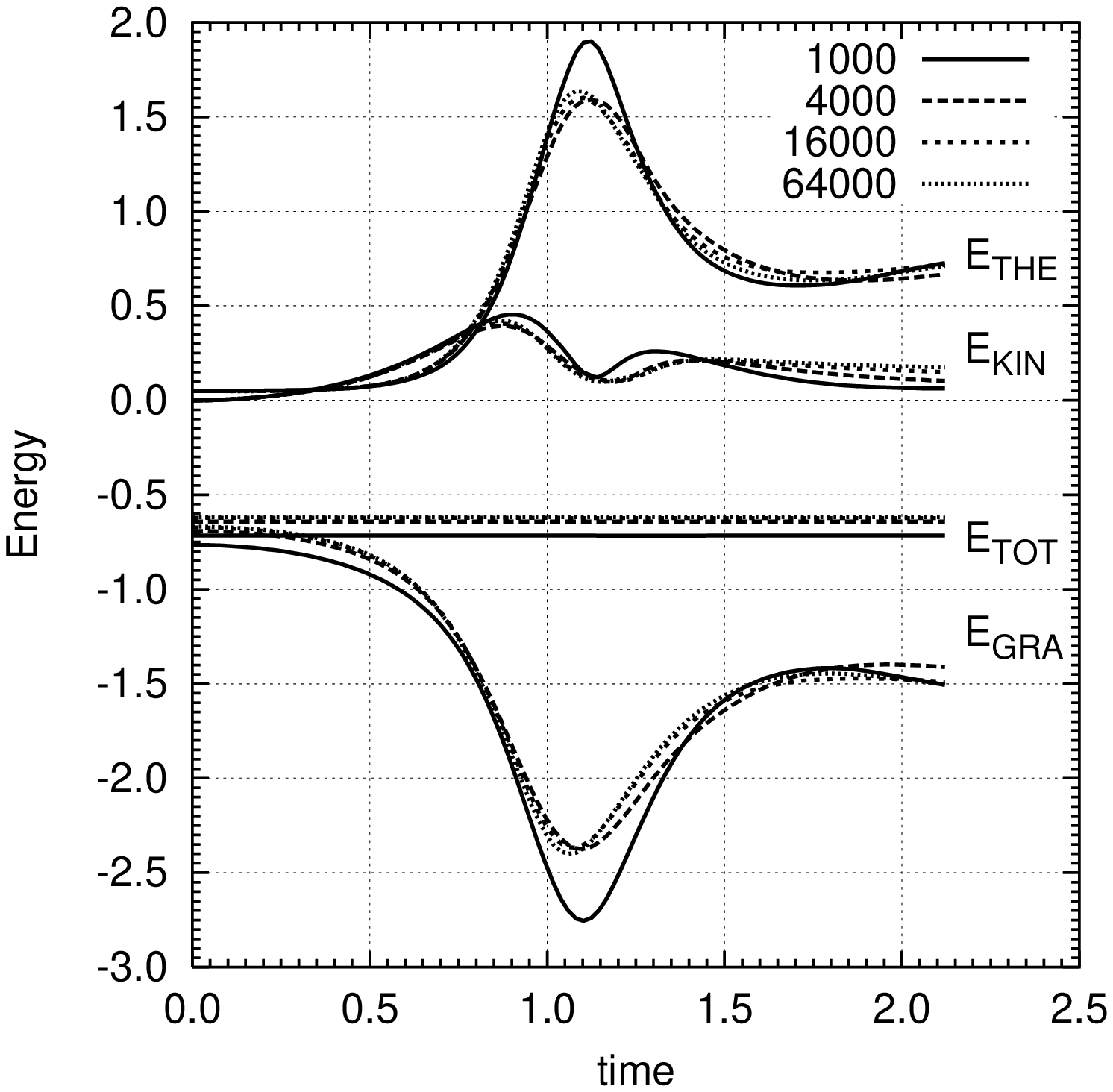} & \includegraphics[width=2.2in]{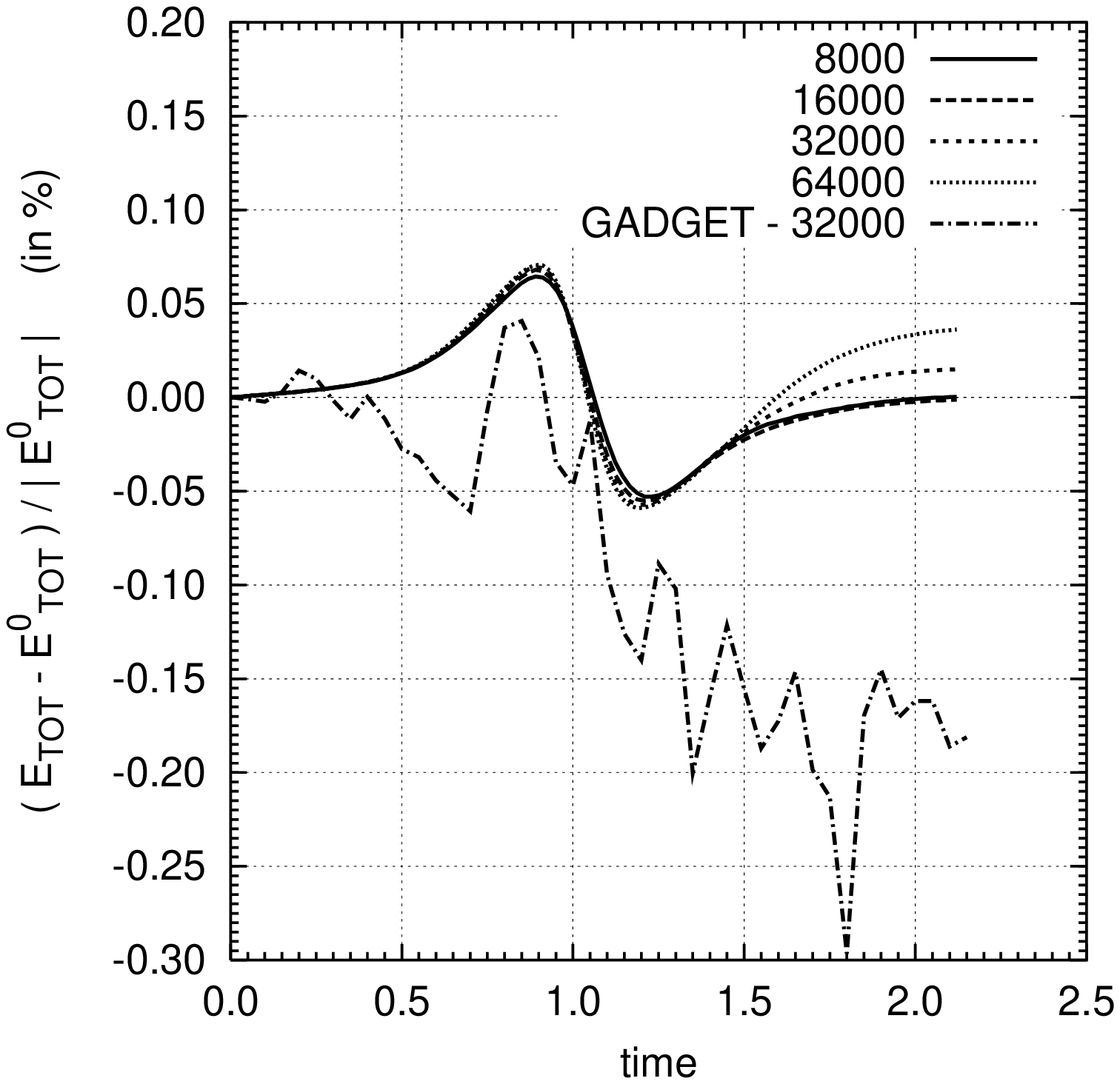}
\end{tabular}}
\caption{Time evolution of the thermal, kinetic, potential and
total energy for the collapse of an initially isothermal gas
sphere (left) and the relative total energy error for different
models (right). The different lines correspond to the different
numbers of gas particles. In the right figure we also present the
energy error of the {\tt \bf GADGET} code with
``standard'' parameters for the 32,000 particles.}
\label{berccontr1&contr2}
\end{figure}

The self-gravitating collapse of an initially isothermal, cool gas
sphere is a common test problem for SPH codes
\cite{E1988, HK1989, SM1993, CLC1998, TTPCT2000, SYW2001}.
Following these authors, we consider a gas sphere of 
total mass M, radius R, and initial density profile $
\rho(r) = \frac{M}{2~\pi~R^2}~\frac{1}{r} $ and with an 
internal energy per unit mass of $ u = 0.05~\frac{G~M}{R}$. 
At the beginning, the gas particles are at rest. 
We use a system of units with G=M=R=1. 
In Fig.~\ref{berccontr1&contr2}. 
the time evolution of the different energies (left) and of the
relative error of the total energy (right) are displayed. 
In the right figure we also
compare results applying the publicly accessible {\tt \bf
GADGET} code \cite{SYW2001} with ``standard'' parameters for the
32,000 particles. During the central bounce around t $\approx$ 1.1
most of the kinetic energy is converted into heat and a strong
shock wave travels outwards. 
The results of Fig.~\ref{berccontr1&contr2}. agree very well with those of
\inlinecite{SM1993} and \inlinecite{SYW2001}. The maximum
relative total energy error is around 0.05~\% even for low
(8,000) particle numbers.

\section{The multi-phase processes}

In our multi-phase gas code we use a two-component gas
description of the ISM \cite{TBH1992, SHT1997}, 
a cold (10$^2$-10$^4$ K) cloudy component treated by sticky
particles \cite{TH1993} and a diffuse hot gas (10$^4$-10$^7$ K) 
represented by SPH particles. 
The cold clumps follow an empirical mass-radius relation 
\cite{L1981, SRBY1987, M1990, IK2000, TH1993} with 
$ h_{\rm cl} \simeq 50 \sqrt{
m_{\rm cl}/(10^6~{\rm M}_{\odot}) }$
pc and experience dissipation due to cloud-cloud 
collisions and drag forces within the hot gas. 
The  mass exchange between ``cold'' and ``hot'' gas phases 
happens basically through condensation and/or evaporation (CE) 
described by \inlinecite{CMcKO1981} and \inlinecite{KTH1998}. 
The basic parameter switching between both processes, $\sigma_0$,
is set to 0.03 according to \inlinecite{CMcKO1981}.

\begin{figure}[h]
\tabcapfont
\centerline{%
\begin{tabular}{c@{\hspace{0.2in}}c}
\includegraphics[width=2.2in]{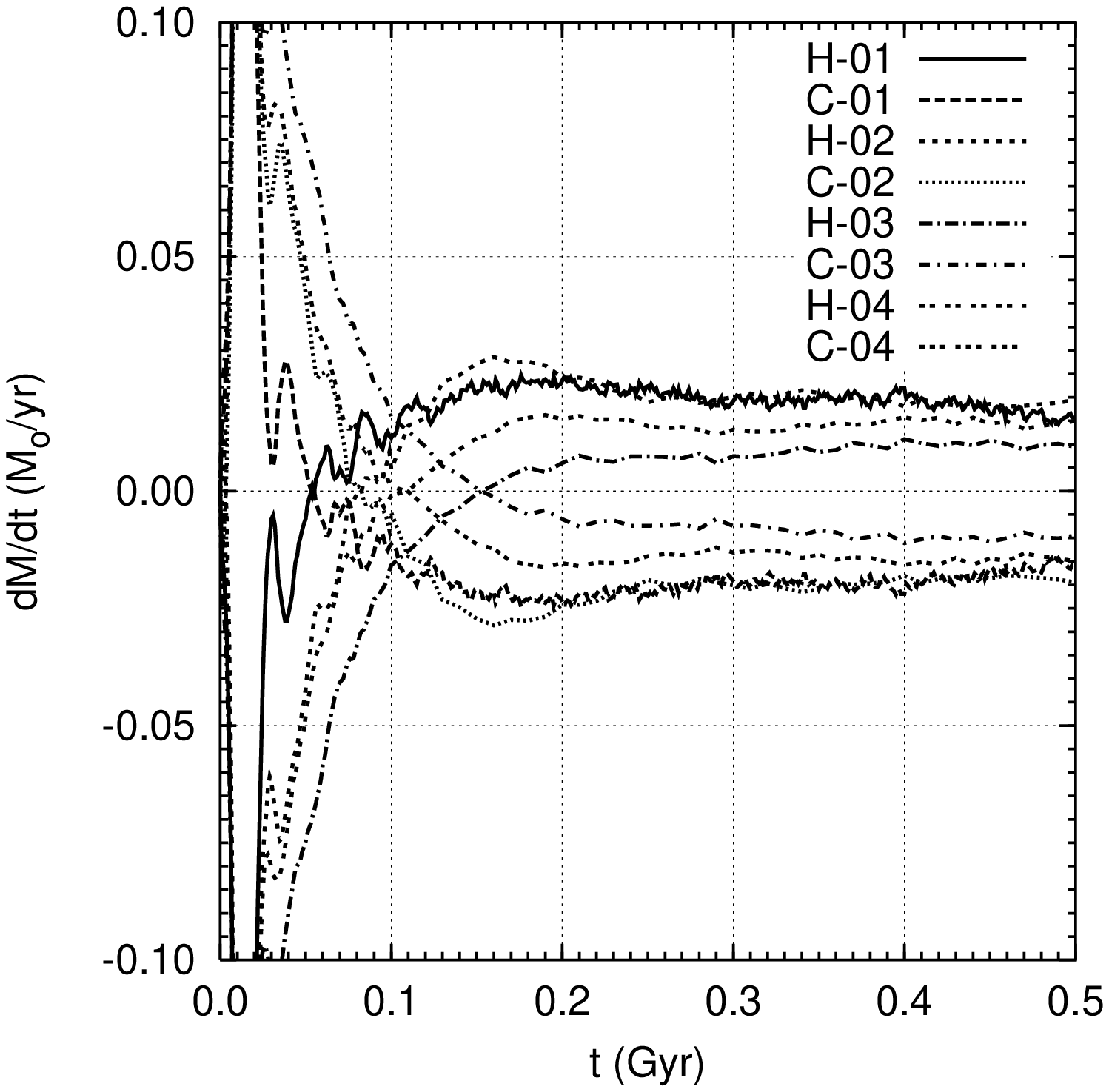} & \includegraphics[width=2.2in]{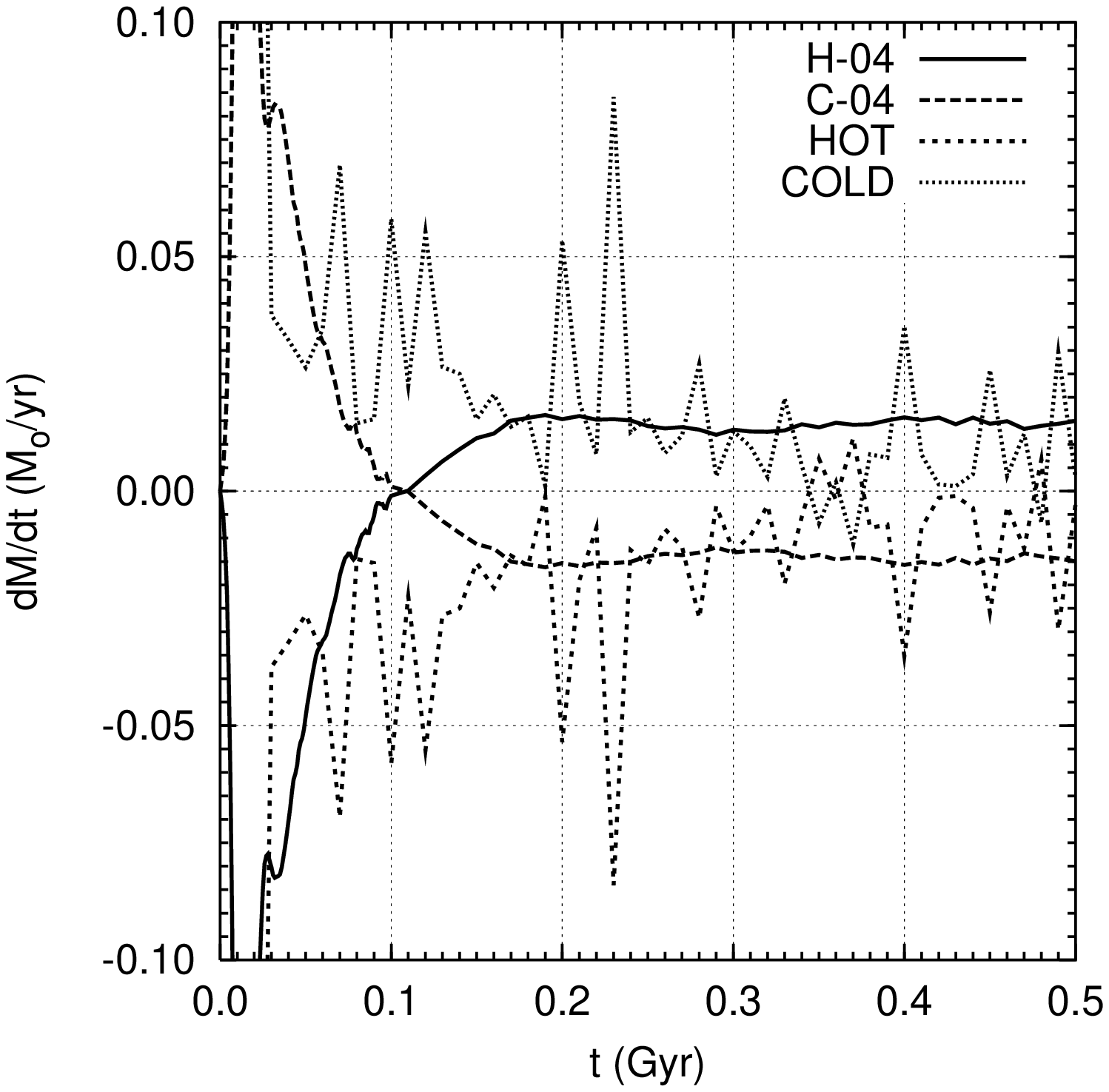}
\end{tabular}}
\caption{The time evolution of the mass exchange rate due to
condensation {\it vs.} evaporation (CE) effects (left). The
letters indicate the phase (i.e. H - Hot; C - Cold) and the
numbers indicate the different models. In the right figure we
compare our ``best'' model with 2d multi phase hydro mesh code
results, for similar initial conditions.}
\label{bercmass1&and1}
\end{figure}

As a CE test for our code, we calculate the evolution 
of a ``hypothetical'' gas system inside a dark matter (DM) 
halo and compare the results with a 2d multi-phase hydro 
mesh-code \cite{RH2000} for the same system. The initial total 
gas mass amounts to $2 \times 10^9$ M$_\odot$ (99\% ``{\tt COLD}'' 
+ 1\% ``{\tt HOT}'') and is distributed according to a Plummer-Kuzmin 
disk with parameters a=0.1~kpc and b=2~kpc \cite{MN1975}.
The initial temperature for the cold gas is set to 2$\times$10$^3$ K, 
for the hot gas to 10$^6$ K. The DM halo is static with parameters 
r$_0$=2~kpc and $\rho_0=2$ M$_\odot$/pc$^3$ \cite{B1995} so that its
mass inside the initial gas distribution (20~kpc) amounts to 
$\simeq 3.2 \times 10^{11}$ M$_\odot$. 

In some of the models we also use an initial cloud mass function 
(ICMF) \cite{MT2000} with a power-law index -1.5 and lower and 
upper mass limits at 10$^4$~M$_\odot$ and 10$^6$~M$_\odot$, 
respectively. 
In Fig.~\ref{bercmass1&and1}. we present the time evolution of the mass
exchange rate via CE for different models: {\tt VER-01} with 
5,000 hot-gas SPH particles of identical mass and 5,000 cold gas 
particles, {\tt VER-02} with 10,000 particles each,  
{\tt VER-03} with the same particle numbers as in {\tt VER-01} but 
an ICMF, and {\tt VER-04} equal to {\tt VER-02} but also with an 
ICMF. The figures show that the basic behaviour of mass exchange 
between the hot and cold phase is well  described with our model
and almost independent of the particle number and ICMF, but in 
agreement on average with the rates of the 2d mesh code.

\vspace{-0.5cm}

\acknowledgements The work was supported by the German Science
Foundation (DFG) under grant {\it 436 UKR} and by the {\it SFB 439} 
at the University of Heidelberg. P.B. is grateful for the
hospitality of the Inst.\ f\"ur Theor.\ Physik und Astrophysik in Kiel 
where most of this work was done. The calculations have been performed 
with the {\tt GRAPE5} system at the Astronomical Data Analysis Center
 of the NAO, Japan.

\end{article}
\end{document}